\documentclass[aps,prd,12point,twocolumn,nofootinbib,showpacs,superscriptaddress]{revtex4-2}
\def\theequation{\arabic{section}.\arabic{equation}}
\usepackage{psfrag}
\usepackage{subfigure}
\usepackage{color}
\usepackage{mathrsfs}
\usepackage{graphicx}
\usepackage{amssymb, bm}
\usepackage{amsmath, amsthm}
\usepackage{epstopdf}
\usepackage{hyperref}
\usepackage{enumerate}
\usepackage{longtable}
\usepackage{amsmath}
\usepackage[normalem]{ulem}

\newcommand{\be}{\begin{equation}}
\newcommand{\ee}{\end{equation}}
\definecolor{pinegreen}{rgb}{0.0, 0.47, 0.44}

\begin{document}
\def\theequation{\arabic{section}.\arabic{equation}}

\title{Stealth metastable state of scalar-tensor thermodynamics}

\author{Valerio Faraoni}
\email[]{vfaraoni@ubishops.ca}
\affiliation{Department of Physics \& Astronomy, Bishop's University, 
2600 College Street, Sherbrooke, Qu\'ebec, Canada J1M~1Z7
}
\author{Th\'eo B. Fran\c{c}onnet}
\email[]{tfranconnet21@ubishops.ca}
\affiliation{Department of Physics \& Astronomy, Bishop's University, 
2600 College Street, Sherbrooke, Qu\'ebec, Canada J1M~1Z7
}
\affiliation{Facult\'e des Sciences, Universit\'e de Montpellier, Place 
Eug\`ene Bataillon, 34090 Montpellier, France}

\begin{abstract} 

We investigate a puzzle in the recent thermodynamics of scalar-tensor 
gravity, in which general relativity is a zero-temperature state of 
equilibrium and scalar-tensor gravity is the non-equilibrium configuration 
of an effective dissipative fluid. A stealth solution of Brans-Dicke 
gravity with constant positive temperature is shown to be analogous to a 
metastable state for the effective fluid and to suffer from an  
instability. The stability analysis employs a 
version of the Bardeen-Ellis-Bruni-Hwang gauge-invariant formalism for 
cosmological perturbations adapted to modified gravity. The metastable 
state is destroyed by tensor perturbations.

\end{abstract}



\maketitle

\section{Introduction}
\label{sec:1}
\setcounter{equation}{0}

Gravity is one of only four fundamental forces known but it behaves 
differently than the electroweak and strong interactions in several 
respects. For this reason, it has been suggested that gravity may not be 
fundamental after all, but rather be emergent, an idea that has been 
formulated in various approaches (see \cite{Sakharov, Visser:2002ew, 
Volovik03, Barcelo:2005fc, Padmanabhan:2008wi,Padmanabhan:2009vy, 
Hu:2009jd, Verlinde:2010hp, Carlip:2012wa, Giusti:2019wdx} for reviews).  
A particularly influential approach was that of Jacobson's thermodynamics 
of spacetime in which the Einstein equation was derived with purely 
thermodynamical 
considerations \cite{Jacobson:1995ab}. Later on, scalar-tensor gravity (in 
its incarnation as metric $f(R)$ gravity) was examined and its field 
equations were again derived from thermodynamics \cite{Eling:2006aw}. 
Moreover, the idea was advanced that general relativity (GR) constitutes a 
state of equilibrium while modified gravity corresponds to an out-of 
equilibrium state (\cite{Eling:2006aw}, see also \cite{Chirco:2010sw}). 
This idea is not outrageous if one thinks that the field content of 
scalar-tensor gravity consists of the two massless spin two modes of GR 
plus a (usually massive) scalar mode propagating as well. Exciting this 
scalar mode corresponds to an excited state with respect to GR.

In spite of a large literature, the order parameter, or equations, 
describing the approach to equilibrium have never been found. Recently, a 
different approach was proposed for scalar-tensor gravity, including  
a definition of ``temperature of gravity'' quantifying the proximity (or 
lack thereof) of a theory of gravity to GR, and an equation describing the 
approach to equilibrium \cite{Faraoni:2018qdr, Faraoni:2021lfc, 
Faraoni:2021jri,Giusti:2021sku, Giardino:2022sdv}.  We refer to this 
approach of \cite{Faraoni:2018qdr, Faraoni:2021lfc, Faraoni:2021jri, 
Giusti:2021sku} as ``thermodynamics of scalar-tensor gravity''  and we stress 
that, in spite of similarities with the basic idea of 
Ref.~\cite{Jacobson:1995ab,Eling:2006aw}, it is {\em not} thermodynamics 
of spacetime (as commonly the work inspired by 
\cite{Jacobson:1995ab,Eling:2006aw} is referred to), but it is a very 
different approach. Although less insightful in fundamental aspects, at 
the same time it is minimalistic in its assumptions, which is the reason 
why it allows progress in finding an effective ``temperature of gravity'' 
and in describing the approach to the GR equilibrium state 
\cite{Faraoni:2018qdr, Faraoni:2021lfc, Faraoni:2021jri, Giusti:2021sku}.  

The key feature of this new formalism is that the field equations of 
scalar-tensor gravity are written as effective Einstein equations by 
moving all terms containing $\phi$ and its first and second 
derivatives to the right-hand side, where they form an effective 
stress-energy tensor $T_{ab}^{(\phi)}$. It is then shown that, when the 
gradient $\nabla^a\phi$ is timelike, $T_{ab}^{(\phi)}$ has the form of the 
stress-energy tensor of a dissipative fluid 
\cite{Pimentel89,Faraoni:2018qdr, Quiros:2019gai, Giusti:2021sku},  to 
which one can apply the 
constitutive 
relations of Eckart's first order thermodynamics \cite{Eckart40}. Within 
the well-known limitations of Eckart's theory ({\em e.g.}, 
\cite{Maartens:1996vi, Andersson:2006nr}), these relations lead to the 
definition and physical interpretation of fluid quantities as effective 
``temperature of gravity'' ${\cal T}$, thermal conductivity ${\cal K}$, 
heat flux density, anisotropic stresses, and shear and bulk viscosity 
coefficients \cite{Faraoni:2018qdr, Faraoni:2021lfc, Faraoni:2021jri}. 
This formalism has since been extended to Horndeski gravity 
\cite{Giusti:2021sku} and applied to Friedmann-Lema\^itre-Robertson-Walker 
(FLRW) cosmology \cite{Giardino:2022sdv} (its application to other 
situations in gravity is in progress).

The thermodynamics of scalar-tensor gravity is consistent (within the 
limitations of Eckart's theory) in both the general theory and its 
application to specific analytical solutions but, in this context, a 
little puzzle remains. There is a stealth solution of Brans-Dicke theory 
that corresponds to a state of constant non-zero temperature 
\cite{Faraoni:2021jri}, which is difficult to interpret. The present paper 
is devoted to solving this conundrum and learning a lesson that is 
important for the general theory beyond specific solutions.  Here we show 
that this Brans-Dicke stealth solution can be interpreted as a sort of 
metastable state of the theory which exists at constant (non-zero) 
temperature and, therefore, always remains far away from the GR state of 
equilibrium (which corresponds to zero temperature instead). However, this 
state 
is unstable and small perturbations destroy it, as it happens for 
supercooled water, which freezes as soon as its  
container is shaken, or impurities acting as ice nucleation nuclei are 
thrown into it. In our case, the perturbations are either scalar or tensor 
perturbations and the stability analysis is performed using 
a gauge-invariant formalism designed for modified gravity.

Let us be more specific about the theory, the solution, and the 
thermodynamical formalism (we follow the notation of 
Ref.~\cite{Waldbook}). 
The context is the Brans-Dicke theory of gravity, described by the action 
\cite{Brans:1961sx}
\be
S_\mathrm{BD} = \int d^4 x \, \sqrt{-g} \left[ \phi R -\frac{\omega}{\phi} 
\, \nabla^c \phi \nabla_c \phi  -V(\phi) \right] \label{eq:1.1}
\ee
in vacuo, where $\phi$ is the Brans-Dicke scalar field corresponding, 
approximately, 
to the inverse of the effective gravitational coupling $G_\mathrm{eff}$, 
$R$ is the Ricci scalar, $\omega$ is a constant (``Brans-Dicke 
coupling''),  $V(\phi)$ is a scalar field potential, and $g$ is the 
determinant of the metric $g_{ab}$. 

Several analytical solutions of Brans-Dicke gravity are known, 
including time-dependent ones (see the review \cite{Faraoni:2021nhi}). 
The  solution of interest here \cite{Faraoni:2017ecj} corresponds to the 
choice
\be
\omega=-1 \,, \quad\quad V(\phi)=V_0 \, \phi \,,
\ee
where $V_0$ is  a positive constant.\footnote{Since 
$\phi=G_\mathrm{eff}^{-1}>0$, the potential is then effectively bounded 
from below.} The line element and scalar field 
in spherical coordinates $\left( t,r, \vartheta, \varphi \right)$ are
\cite{Faraoni:2017ecj}
\be
ds^2 =-dt^2 +A^{-\sqrt{2}}(r) \, dr^2 
+ A^{1-\sqrt{2}} (r) \, r^2 d\Omega_{(2)}^2 \,,
\ee

\be
\phi(t,r) =\phi_0 \, \mbox{e}^{2a_0 t}  A^{1/\sqrt{2}}(r) \,,
\ee
where $A(r) =1-2m/r$, $ m, a_0, \phi_0$ are constants, and 
$d\Omega_{(2)}^2 = d\vartheta^2 + \sin^2 \vartheta \, d\varphi^2$ is the 
line element on the unit 2-sphere. This geometry is conformal to that of 
the Fonarev solution of GR \cite{Fonarev:1994xq} and can be seen as a 
special case of the Campanelli-Lousto family of solutions 
\cite{CampanelliLousto, CampanelliLousto2}, but the functional form of the 
scalar field is different from the Campanelli-Lousto one 
\cite{CampanelliLousto, CampanelliLousto2}. Since $\omega=-1$ Brans-Dicke 
gravity is the low-energy limit of bosonic string theory 
\cite{Callan:1985ia,Fradkin:1985ys}, presumably there is some stringy 
analogue of this solution. Here, however, we are only interested in the 
special case obtained by the limit $m\rightarrow 0$, which produces the 
Minkowski metric
\be
ds^2 =-dt^2 +dr^2 +r^2 d\Omega_{(2)}^2 \label{stealth1}
\ee
with
\be
\phi(t)= \phi_0 \, \mbox{e}^{2a_0t} \,,\label{stealth2}
\ee
a little-known stealth solution in which the scalar field does not 
gravitate. In Ref.~\cite{Faraoni:2021jri}, this analytical solution was 
examined as an example of the new thermodynamics of scalar-tensor gravity, 
resulting in a surprise:  in general, the product of thermal conductivity 
and temperature is given by \cite{Faraoni:2021lfc, Faraoni:2021jri}
\be
{\cal KT}= \frac{ \sqrt{ -\nabla^c\phi \nabla_c\phi}}{8\pi \phi} 
\ee
which, for the stealth solution~(\ref{stealth1}), (\ref{stealth2}) 
becomes constant \cite{Faraoni:2021jri}, 
\be
{\cal KT}= \frac{|a_0|}{4\pi} \,.
\ee
Such states are in principle possible because the approach to the GR 
equilibrium is described by the equation \cite{Faraoni:2021jri}
\be
\frac{d \left( {\cal KT}\right)}{d\tau} = 8\pi \left( {\cal KT}\right)^2 
-\Theta {\cal  KT} +\frac{\Box \phi}{8\pi \phi} \,,
\ee
where $\tau$ is the proper time of the effective $\phi$-fluid, its 
4-velocity is
\be
u^a=\frac{\nabla^a\phi}{\sqrt{ - \nabla^c\phi \nabla_c\phi}} \,,
\ee
and $\Theta$ is its expansion scalar, given by \cite{Faraoni:2018qdr}
\be
\Theta = \nabla _a u^a 
= \frac{1}{\sqrt{-\nabla_c \phi \nabla ^c \phi}} 
\left( \Box \phi - \frac{\nabla ^a \phi \nabla ^b \phi \nabla _a \nabla 
_b \phi}{\nabla _e \phi \nabla ^e \phi} \right) \,.\label{expansion}
\ee 
In order for $u^a$  to be 
future-oriented for the specific solution~(\ref{stealth1}), 
(\ref{stealth2}), it must be $a_0<0$ \cite{Faraoni:2021jri}. It clear that 
states ${\cal KT}=$~const.$\equiv C$ can be obtained if 
\be
C^2 -C \Theta+\frac{\Box \phi}{8\pi \phi}=0 \,.\label{short}
\ee
For the particular solution~(\ref{stealth1}), (\ref{stealth2}) it is 
$\Theta=0$, while the equation of motion of the Brans-Dicke field
\be
\Box\phi = \frac{1}{2\omega+3} \left( \phi\, \frac{dV}{d\phi} -2V \right) 
\ee
gives $\Box\phi= -V_0\, \phi$ for $\omega=-1$, $V(\phi) =V_0 \, \phi$, and 
$\phi=\phi_0 \, \mbox{e}^{2a_0 \phi}$. As a result, Eq.~(\ref{short}) is 
satisfied if $V_0=4a_0^2$, a relation derived also in 
\cite{Faraoni:2021jri} 
with indipendent considerations.

The physical interpretation of this state with constant ${\cal KT}> 0$   
remained unclear and is the 
subject of the present paper. We show that this solution can be 
interpreted as  a sort of metastable state of the theory which exists at 
constant (non-zero) temperature, and therefore always remains far away 
from the GR state of equilibrium (which corresponds to ${\cal KT}=0$). 
However this state is unstable with respect to  
tensor perturbations of the 
spacetime~(\ref{stealth1})~(\ref{stealth2}). The stability analysis with 
respect to these perturbations is carried out using  
the Bardeen-Ellis-Bruni-Hwang gauge-invariant formalism developed for 
cosmological perturbations in \cite{Bardeen, EllisBruni89, 
EllisHwangBruni89, EllisBruniHwang90, 
HwangVishniac90} and adapted to modified gravity in 
\cite{Hwang90, Hwang90b, Hwang:1995bv, Hwang:1996bc, 
Hwang:1996xh, Hwang97, Noh:2001ia}. It 
can be applied because the stealth Minkowski spacetime~(\ref{stealth1}), 
(\ref{stealth2}) is a trivial FLRW spacetime. 

The instability matches the intuition that this stealth state of gravity  
is analogous to a metastable state in which 
a fluid (in our case, an effective fluid) can remain hanged for a while, 
but is destroyed by arbitrarily small perturbations. Sec.~\ref{sec:2} 
recalls the needed equations of the gauge-invariant formalism for 
cosmological perturbations of scalar-tensor gravity, which are solved in 
Sec.~\ref{sec:3}, establishing the presence of instability. 
Sec.~\ref{sec:4} contains conclusions on the significance of this 
metastable state for the thermodynamics of scalar-tensor gravity.

\section{Equations of gauge-invariant perturbation theory for modified 
gravity}
\label{sec:2}
\setcounter{equation}{0}

Gauge-invariant cosmological perturbation theory in modified gravity 
was developed in 
\cite{Hwang90, Hwang90b, Hwang:1995bv, Hwang:1996bc, Hwang:1996xh, 
Hwang97, Noh:2001ia} for a rather general class of theories, {\em 
i.e.}, mixed  scalar-tensor/$f\left(R\right)$ 
gravity in the metric formalism. The vacuum action is 
\begin{equation}
S=\int  d^{4}x\,\sqrt{-g}\,\left[\frac{f(\phi,R)}{2}
-\frac{ \bar{\omega}(\phi)}{2} \, \nabla^c \phi\nabla_c \phi
-V\left(\phi\right)\right].\label{eq:2.1}
\end{equation}
Assuming a spatially flat  FLRW  universe with line element
\be \label{FLRWmetric}
ds^2=-dt^2+a^2(t) \left( dx^2+dy^2 +dz^2 \right)
\ee
the corresponding field equations analogous to the Einstein-Friedmann 
equations of GR read 
\begin{eqnarray}
H^{2} & = & \frac{1}{3F} \left(\frac{ \bar{\omega}}{2} \, \dot{\phi}^{2} 
+\frac{RF}{2}-\frac{f}{2}+V-3H\dot{F}\right)\,\,,
\label{eq:12} \\
&&\nonumber\\
\dot{H} & = & -\frac{1}{2F}\left( \bar{\omega} \, \dot{\phi}^{2} 
+\ddot{F}-H\dot{F}\right)\,\,,\label{eq:13}\\
&&\nonumber\\
\ddot{\phi} & + & 3H\dot{\phi}+\frac{1}{2 \bar{\omega}} 
\left(\frac{d \bar{\omega}}{d\phi} \, \dot{\phi}^{2}-\frac{\partial f}{ 
\partial\phi}+2 \, \frac{dV}{d\phi}\right)=0\,\,,\label{eq:14}
\end{eqnarray}
where an overdot denotes differentiation with respect to the comoving 
time $t$,  $H\equiv\dot{a}/a$ is the Hubble function, and $F\equiv\partial 
f/\partial R$. The metric perturbations in the 
Bardeen-Ellis-Bruni-Hwang formalism 
\cite{Bardeen, EllisBruni89, EllisHwangBruni89, EllisBruniHwang90, 
HwangVishniac90} are completely described by the quantities  $A,\, 
B,\, H_{L}$, and $H_{T}$ defined by
\begin{eqnarray}
g_{00} & = & -a^{2}\left(1+2AY\right)\,\,,\label{eq:15}\\
&&\nonumber\\
g_{0i} & = & -a^{2}BY_{i}\,\,,\label{eq:16} \\
&&\nonumber\\
g_{ij} & = & a^{2}\left[h_{ij}\left(1+2H_{L}\right) 
+2H_{T}Y_{ij}\right]\,,\label{eq:17}
\end{eqnarray}
where $h_{ij}$ is the three-dimensional metric of the FLRW background,  
the scalar harmonics $Y$ are the eigenfunctions of the eigenvalue problem 
$\bar{\nabla}_{i}\bar{\nabla}^i Y=-k^{2}Y$, $k$ is the corresponding  
eigenvalue,  and $\bar{\nabla_{i}}$ is the covariant derivative operator 
of  $h_{ij}$.  The  vector and tensor  harmonics $Y_{i}$ and $Y_{ij}$ 
are given by 
\begin{eqnarray}
Y_{i} &=& -\frac{1}{k} \, \bar{\nabla}_{i}Y\,\,,\label{eq:18}\\
&&\nonumber\\
Y_{ij} &=& \frac{1}{k^2} \, \bar{\nabla}_{i} 
\bar{\nabla}_{j}Y+\frac{1}{3}Yh_{ij}\,\,,\label{eq:19}
\end{eqnarray}\\
respectively. The Bardeen gauge-invariant potentials 
\cite{Bardeen} are
\begin{eqnarray}
\Phi_{H} & = & H_{L}+\frac{H_{T}}{3} +\frac{\dot{a}}{k} \left(B 
-\frac{a}{k}\,\dot{H}_{T}\right)\,,\label{eq:20}\\
&&\nonumber\\
\Phi_{A} & = & A+\frac{\dot{a}}{k}\left(B 
-\frac{a}{k}\,\dot{H}_{T}\right) 
+\frac{a}{k}\left[\dot{B}-\frac{1}{k} 
\left(a\dot{H}_{T}\right)\dot{} \,\right]\,,\nonumber\\
&&  \label{eq:21}
\end{eqnarray}
the Ellis-Bruni variable \cite{EllisBruni89} is
\begin{equation}
\Delta \phi=\delta\phi+\frac{a}{k} \, \dot{\phi}\left(B 
-\frac{a}{k} \, \dot{H}_{T}\right)\,\,,\label{eq:22}
\end{equation}
while similar relations define the gauge-invariant variables 
$\Delta f,\,\Delta F,$ and $\Delta R$. 

To first order, the gauge-invariant perturbations satisfy the (redundant) 
system of equations \cite{Hwang90, Hwang90b, Hwang:1995bv, Hwang:1996bc, 
Hwang:1996xh, Hwang97, Noh:2001ia}
\begin{widetext}
\begin{eqnarray} 
\label{26} && \Delta \ddot{\phi} + \left( 3H + 
\frac{\dot{\phi}}{ \bar{\omega}} \, \frac{d \bar{\omega}}{d\phi} \right) 
\Delta \dot{\phi} + \left[ \frac{k^2}{a^2}
 + \frac{\dot{\phi}^2}{2} \frac{d}{d\phi} \left( 
\frac{1}{\bar{\omega}} \frac{d\bar{\omega}}{d\phi} \right) -\, \frac{d}{ 
d\phi} \left( \frac{1}{2\bar{\omega}} \frac{\partial f}{\partial \phi} 
-\frac{1}{\bar{\omega}} \frac{dV}{d \phi} \right) \right] \Delta \phi 
\nonumber \\ && \nonumber \\ && =
  \dot{\phi} \left( \dot{\Phi}_A - 3\dot{\Phi}_H \right)
 + \frac{\Phi_A}{\bar{\omega}} \left( \frac{\partial f}{\partial \phi} 
-2 \, \frac{dV}{d\phi} \right)
 +\frac{1}{2\bar{\omega}} \, \frac{\partial^2 f}{\partial \phi 
\partial R} \, \Delta R \, , \end{eqnarray}

\begin{eqnarray} \label{eq:27} && \Delta \ddot{F} +3H \Delta 
\dot{F} +\left( \frac{k^2}{a^2} - \frac{R}{3} \right) \Delta F 
+\frac{F}{3} \, \Delta R + \frac{2}{3} \, \bar{\omega} \dot{\phi} 
\Delta \dot{\phi} +\frac{1}{3} \left( \dot{\phi}^2 
\frac{d\bar{\omega}}{d\phi} + 2\frac{\partial f}{\partial \phi} -4 \, 
\frac{dV}{d\phi} \right) \Delta \phi \nonumber \\ && \nonumber 
\\ && = \dot{F} \left( \dot{\Phi}_A - 3\dot{\Phi}_H \right) + 
\frac{2}{3} \left( FR -2f +4V \right)  \Phi_A \, , 
\end{eqnarray}

\be \label{eq:28} \ddot{H}_T +\left( 3H+ \frac{\dot{F}}{F} 
\right) 
\dot{H}_T +\frac{k^2}{a^2} \, H_T=0 \,, 
\ee

\be \label{eq:29} - \dot{\Phi}_H +\left( H + \frac{\dot{F}}{2F} 
\right) \Phi_A = \frac{1}{2} \left( \frac{ \Delta \dot{F} }{F} 
-H \frac{ \Delta F}{F} + \frac{ \bar{\omega}}{F} \, \dot{\phi} \, 
\Delta \phi \right)  \, , \ee

\begin{eqnarray} \label{eq:30} & & \left( \frac{k}{a} \right)^2 
\Phi_H +\frac{1}{2} \left( \frac{ \bar{\omega}}{F } \dot{\phi}^2 + 
\frac{3}{2} \frac{\dot{F}^2}{F^2} \right) \Phi_A = \frac{1}{2} 
\left\{ \frac{3}{2} \frac{ \dot{F} \Delta \dot{F} }{F^2} + 
\left( 3\dot{H} - \frac{k^2}{a^2} -\frac{3H}{2} \frac{ 
\dot{F}}{F} \right)  \frac{ \Delta F}{F} \right. \nonumber \\ && 
\nonumber \\ & & \left. +\frac{\bar{\omega}}{F} \dot{\phi} \Delta 
\dot{\phi} + \frac{1}{2F} \left[ \dot{\phi}^2 
\frac{d\bar{\omega}}{d\phi} -\frac{ \partial f}{\partial \phi} 
+2\frac{dV}{d\phi} +6\bar{\omega} \dot{\phi} \left( H + \frac{ \dot{F} 
}{2F} \right) \right] \Delta \phi \right\} \, , \end{eqnarray}

\be \label{eq:31} 
\Phi_A + \Phi_H = - \frac{\Delta F }{F} \, , 
\ee

\begin{eqnarray} \label{eq:32}
& & \ddot{\Phi}_H + H \dot{\Phi}_H + \left( H 
+ 
\frac{ \dot{F}}{2F} \right) \left( 2\dot{\Phi}_H -\dot{\Phi}_A 
\right) +\frac{ 1 }{2F} \left( f-2V -RF \right)  \Phi_A 
\nonumber \\ && \nonumber \\ & & = - \frac{1}{2} \left[ \frac{ 
\Delta \ddot{F}}{F} + 2H \, \frac{\Delta \dot{F}}{F} + \left( 
P-\rho \right)  \frac{ \Delta F}{2F} + \frac{ \bar{\omega}}{F} \, 
\dot{\phi} \, \Delta \dot{\phi} + \frac{1}{2F} \left( 
\dot{\phi}^2 \, \frac{d\bar{\omega}}{ d\phi} +\frac{\partial 
f}{\partial \phi } -2 \, \frac{dV}{ d\phi } \right) \Delta \phi 
\right] \, , \nonumber \\ 
&& 
\end{eqnarray} and 
\be \Delta R=6 
\left[ \ddot{\Phi}_H +4H\dot{\Phi}_H +\frac{2}{3} 
\frac{k^2}{a^2} \Phi_H -H\dot{\Phi}_A -\left( 2\dot{H}+4H^2 
-\frac{k^2}{3a^2} \right) \Phi_A \right] \,.  \label{eq:33} 
\ee 
\end{widetext}
Although this system is complicated, it simplifies substantially in the 
case of the Minkowski background associated with the analytical solution 
of Brans-Dicke gravity under examination with homogeneous, but 
time-dependent, stealth scalar 
field.  Even with this simplification, solving these equations is 
non-trivial.

\section{Stability of the constant ${\cal KT}$ stealth solution}
\label{sec:3}
\setcounter{equation}{0}

Let examine the equations for gauge-invariant perturbations and assess the    
stability of the stealth solution~(\ref{stealth1}), 
(\ref{stealth2}). We set 
\be
H=0\,, \quad \dot{H}=0\,, \quad a=1\,, \quad\,R=0 \,,
\ee
then the comparison of the actions~(\ref{eq:1.1}) and (\ref{eq:2.1}) 
with $\omega=-1$ and $V=V_0 \, \phi$ yields
\be
f(\phi, R  )=2\phi\,R, \quad\quad 
\bar{\omega}(\phi)=-\frac{2}{\phi}\,, 
\quad\quad F=2\phi \,,
\ee
while Eq.~(\ref{stealth2}) gives  
$ \dot{\phi}/\phi  = 2a_0$.

The gauge-invariant equations listed in Sec.~\ref{sec:2} simplify 
considerably. Equation(\ref{eq:28}) for the tensor modes decouples from 
the other equations, assuming the form  
\be \label{grav_wave}
\ddot{H}_T+ 2a_0\dot{H}_T+k^2H_T=0 \,.
\ee
The term containing $\dot{H}_T$ describes friction if $a_0>0$ and 
anti-friction if $a_0<0$, therefore, tensor modes are stable if 
$a_0\geqslant0$ and unstable if $a_0<0$. As seen in Sec.~\ref{sec:2}, it 
must be $a_0<0$ in order for the four-velocity of the effective 
$\phi$-fluid to be future-oriented, and we conclude that the stealth 
solution is unstable with respect to tensor modes of short wavelengths.

Proceeding, Eqs.~(\ref{eq:29})-(\ref{eq:31}) give
\be \label{new29}
-\dot{\Phi}_H+ a_0\Phi_A = \frac{1}{2}\left(\frac{\Delta\dot{\phi}}{\phi} 
-2a_0 \frac{\Delta\phi}{\phi}\right) \,,
\ee
\begin{eqnarray} 
\label{new30} 
& & k^2\Phi_H + {a_0}^2\Phi_A \nonumber\\ && \nonumber\\ 
& & = \frac{1}{2}\left[a_0\frac{\Delta\dot{\phi}}{\phi} + 
\frac{\Delta\phi}{\phi}\left(-k^2 + \frac{V_0 -8{a_0}^2}{2}\right)\right] 
\,, 
\end{eqnarray}
and
\be \label{new31}
\Phi_A + \Phi_H = -\frac{\Delta\phi}{\phi} \,.
\ee
Equation~(\ref{eq:33}) gives
\be \label{new33}
\Delta R = 6\left(\ddot{\Phi}_H + \frac{2k^2}{3} \, \Phi_H + 
\frac{k^2}{3} \, \Phi_A\right) 
\ee
which, substituted into Eq.~(\ref{eq:27}), yields
\begin{eqnarray}  
& & \Delta\ddot{\phi}-\frac{4a_0}{3} \, \Delta\dot{\phi} +\left(k^2 + 
\frac{4}{3}{a_0}^2 -\frac{2}{3}V_0\right)\Delta\phi \nonumber\\ 
&& \nonumber\\ 
& & = \phi\left[-2\ddot{\Phi}_H + 2a_0\left(\dot{\Phi}_A - 
3\dot{\Phi}_H\right) \right. \nonumber \\ 
&&\nonumber\\ 
& & \left. + \frac{4}{3} \left(V_0-\frac{k^2}{2}\right)\Phi_A 
- \frac{4k^2}{3} \, \Phi_H\right] \,.\label{new27}
\end{eqnarray}
Using Eq.~(\ref{new31}), we can now eliminate $\ddot{\Phi}_A$, 
$\dot{\Phi}_A$, and $\Phi_A$;  Eqs.~(\ref{new29}) and 
(\ref{new30}) simplify to 
\be
\dot{\Phi}_A +a_0\Phi_A + \frac{\Delta\dot{\phi}}{2\phi} - 
a_0 \, \frac{\Delta\phi}{\phi} = 0 \,, \label{eq:3.9}
\ee

\be
\left(k^2-{a_0}^2\right)\Phi_A = -\frac{a_0}{2} \, \frac{\Delta 
\dot{\phi} }{\phi} - 
\left(\frac{k^2}{2} +\frac{V_0}{4}-2a_0^2 \right) \frac{\Delta\phi}{\phi} 
\ee
which, in turn, allows one to eliminate $\ddot{\Phi}_A$, $\dot{\Phi}_A$,  
and $\Phi_A$. Equation~(\ref{new27}) then becomes
\begin{eqnarray} 
&& 
\left(4{a_0}^2 -  V_0\right) \Delta\dot{\phi} 
+ \frac{1}{2a_0} \left[\left(2{a_0}^2-\frac{V_0}{2}\right)k^2 \nonumber 
\right. \nonumber\\ 
&&\nonumber\\ 
& & \left.  + a_0^2 \left(-26{a_0}^2 + \frac{21}{2} \, V_0\right) + 
 V_0^2\right] \Delta\phi = 0 \,.  \nonumber \\ 
&& \label{final27}
\end{eqnarray}
The relation $ V_0=4_0^2 $ of the unperturbed Minkowski space then  
implies that $ \Delta \phi=0 $. However, the Bardeen potentials 
$\Phi_{A,H}$ diverge: in fact, Eq.~(\ref{eq:3.9}) yields
\be
\dot{\Phi}_A=-a_0 \Phi_A \,,
\ee
with solution $\Phi_A(t)=\left( \Phi_A\right)_0 \, \mbox{e}^{-a_0 t} $ 
which diverges as $t\to + \infty$ since $a_0<0$, as already established. 
Equation~(\ref{new31}) gives  the other Bardeen potential 
$\Phi_H=-\Phi_A$, which diverges as well, together with 
\be
\Delta R = -6 \left( a_0^2 +\frac{k^2}{3} \right) \Phi_A \,,
\ee
which follows from Eq.~(\ref{new33}).

\medskip
\section{Conclusions}
\label{sec:4}
\setcounter{equation}{0}

We have uncovered an instability of the stealth solution~(\ref{stealth1}), 
(\ref{stealth2})  with respect to  tensor 
perturbations. The  timescale for the instability 
is $|a_0|^{-1}$.  The only exception occurs for the parameter 
value $a_0=0$ which corresponds to 
Einstein theory, to constant scalar field $\phi=\phi_0 \, \mbox{e}^{2a_0\, 
t}$, and to stability, in agreement with the fact that GR is the 
zero-temperature state of equilibrium for scalar-tensor gravity in the 
thermodynamical formalism \cite{Faraoni:2018qdr, Faraoni:2021lfc, 
Faraoni:2021jri,Giusti:2021sku}.

The information that $a_0$ is negative is crucial to establish the 
instability of the solution~(\ref{stealth1}), (\ref{stealth2}) and comes 
from the requirement that the four-velocity of the effective $\phi$-fluid 
be future-oriented. The correct time orientation is crucial when 
discussing dissipative phenomena, that are irreversible, but no hint  
on the sign of the parameter $a_0$ would be available without the point of 
view of the thermodynamics of scalar-tensor gravity, and the conclusion 
that the 
Brans-Dicke stealth spacetime~(\ref{stealth1}), (\ref{stealth2}) is 
unstable depends crucially on this argument, and only acquires physical 
meaning in this context. Perhaps the only other indication corroborating 
this point is that, if $a_0>0$, then the effective gravitational coupling
$G_\mathrm{eff}=\phi_0^{-1} \, \mbox{e}^{2|a_0|t}$ diverges in the far 
future, which could by itself be taken as signalling instability.  
The lesson to be learned here is that the Brans-Dicke stealth 
spacetime~(\ref{stealth1}), (\ref{stealth2}), describing the analogue of a 
metastable state of the effective $\phi$-fluid with constant ${\cal KT}$, 
is unstable with respect to tensor perturbations. 
In 
retrospect this conclusion could have been expected, but first one needed 
to understand the role of this analytical solution. It is significant that 
the thermodynamics of scalar-tensor gravity gives new meaning to phenomena 
and analytical solutions that would otherwise be unremarkable.

Apart from the specific solution~(\ref{stealth1}), (\ref{stealth2}) and 
the specific $\omega=-1$ Brans-Dicke theory considered (which is, however, 
somehow special as it is the low-energy limit of the bosonic string 
\cite{Callan:1985ia,Fradkin:1985ys}), one learns that non-trivial states 
can in principle exist in the thermodynamics of scalar-tensor gravity. 
Future work will search for other possible metastable states and, above 
all, will check further the consistency of the formalism and its 
consequences for various gravity regimes and physical situations.

Going beyond the particular solution examined here, is known that a 
variety of stealth solutions are possible in Horndeski and in higher-order 
scalar-tensor theories \cite{ Motohashi:2018wdq,Takahashi:2020hso}, a much 
more general framework than Brans-Dicke or ``first generation'' 
scalar-tensor gravity. Presumably, the stability of these stealth 
solutions depends on the details of the scalar fields appearing in them, 
as is the case for the situation examined in the present manuscript. The 
extension of our discussion to these theories is not trivial because the 
Bardeen-Ellis-Bruni-Hwang formalism does not apply directly to 
cosmological perturbations in these theories. Second, the thermodynamics 
of scalar-tensor gravity that motivates this work has been extended to 
``viable'' Horndeski gravities (which turn out to be those that admit an 
Einstein-frame representation) \cite{Giusti:2021sku}, but not to other 
Horndeski and higher order theories. Assessing the stability of stealth 
solutions in these more general scalar-tensor theories will be the 
subject of future research. 

\begin{acknowledgments}

We thank a referee for pointing out an error in a previous version 
of the manuscript and for suggestions improving the presentation. This 
work is supported, in part, by the Natural Sciences \& Engineering 
Research Council of Canada (grant no. 2016-03803 to V.F.).

\end{acknowledgments}


\end{document}